\documentclass[12pt,epsf]{article}
\usepackage{graphicx, amsmath}

\begin{document}

\sloppy
\begin{flushright}{SIT-HEP/TM-29}
\end{flushright}
\vskip 1.5 truecm
\centerline{\large{\bf Dark matter production from cosmic necklaces}}

\vskip .75 truecm
\centerline{\bf Tomohiro Matsuda
\footnote{matsuda@sit.ac.jp}}
\vskip .4 truecm
\centerline {\it Laboratory of Physics, Saitama Institute of
 Technology,}
\centerline {\it Fusaiji, Okabe-machi, Saitama 369-0293, 
Japan}
\vskip 1. truecm
\makeatletter
\@addtoreset{equation}{section}
\def\theequation{\thesection.\arabic{equation}}
\makeatother
\vskip 1. truecm

\begin{abstract}
\hspace*{\parindent}
Cosmic strings have gained a great interest, since they are
formed in a large class of brane inflationary models.
The most interesting story is that cosmic strings in brane
models are distinguished in future cosmological observations. 
If the strings in brane models are branes or superstrings
that can move along compactified space, and also if there are
 degenerated vacua along the compactified space, 
kinks interpolate between degenerated vacua become ``beads'' on the
 strings.
In this case, strings turn into necklaces.
In the case that the compact manifold in not simply connected, 
a string loop that winds around a nontrivial circle is
stable due to the topological reason.
Since the existence of the (quasi-)degenerated vacua and the 
nontrivial circle is a common feature of the brane models, it is
important to study cosmological constraints on the cosmic necklaces and
the stable winding states.
In this paper, we consider dark matter production from
loops of the cosmic necklaces.
Our result suggests that necklaces can put stringent bound on certain
kinds of brane models.
\end{abstract}

\newpage
\section{Introduction}
\begin{figure}[ht]
 \begin{center}
\begin{picture}(400,420)(0,0)
\resizebox{20cm}{!}{\includegraphics{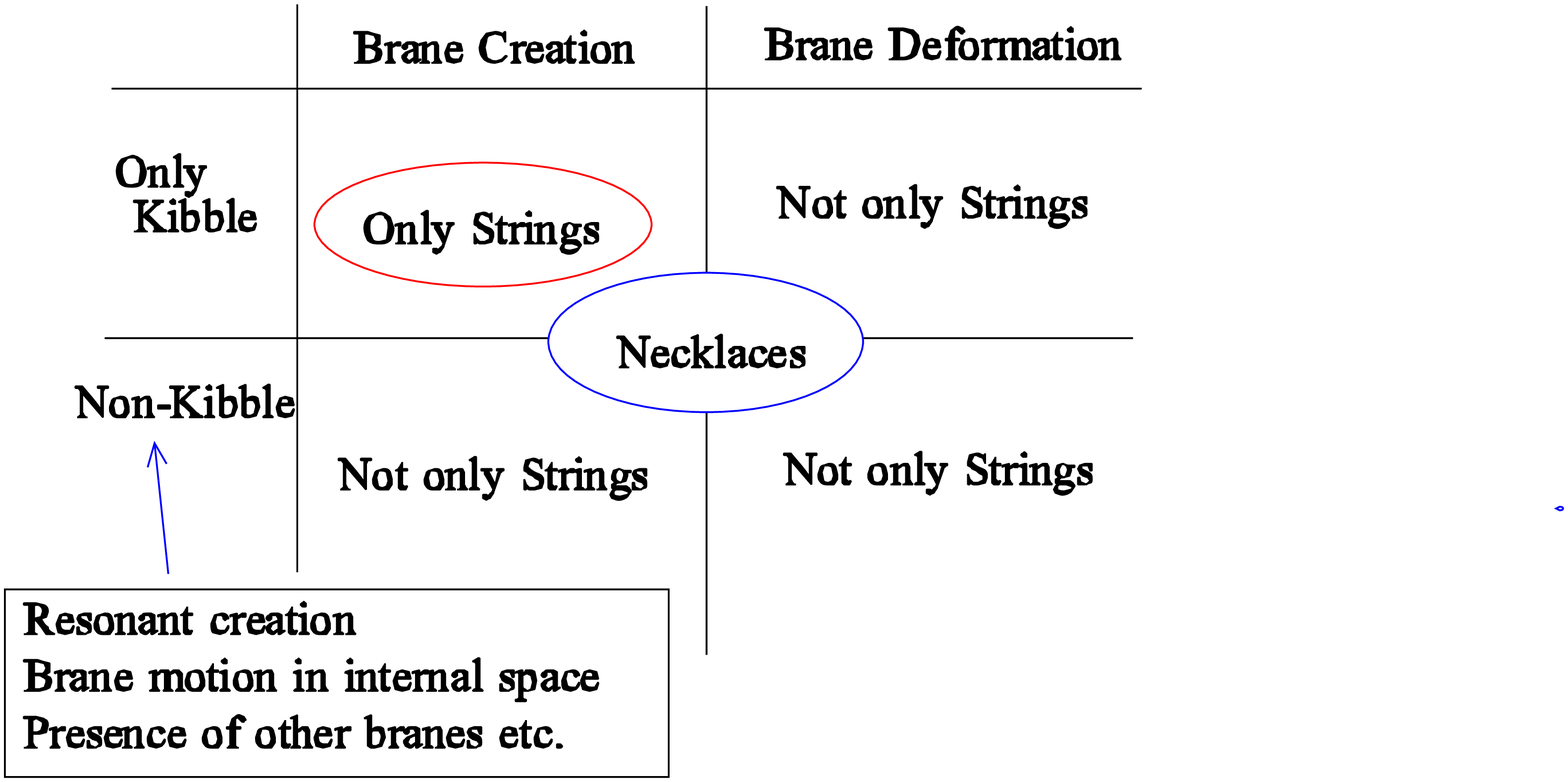}} 
\end{picture}
\caption{If one considers only the conventional Kibble mechanism and the
  brane creation, the resultant cosmological defect should be cosmic
  strings.
  However, the branes (cosmic strings) may move along the direction of
  the internal space and may have kinks on their worldvolume, which look
  like  ``beads'' on the strings.
  Then the strings turn into necklaces, which are the hybrid of the
  brane creation and the brane deformation.}
\label{fig:history}
 \end{center}
\end{figure}

Cosmic strings have gained a great interest, since they are
formed in a large class of brane inflationary models.
In the context of the brane world scenario, cosmic strings are produced
just after brane inflation\cite{brane-inflation0, angled-inflation}.
It has been discussed that such strings
lead to observational predictions that can be used to distinguish 
brane world from conventional phenomenological models\cite{vilenkin-lowp,
matsuda_necklace, matsuda_pbh}.
From phenomenological viewpoints, the idea of large extra
dimension\cite{Extra_1} is important for higher-dimensional models, 
because it may solve or weaken the hierarchy problem. 
In the scenarios of large extra dimension, the fields in the standard
model are localized on wall-like structures, while
the graviton propagates in the bulk. 
In the context of string theory, a natural embedding of
this picture is realized by brane construction.
The brane models are therefore interesting from both the phenomenological
and the cosmological viewpoints.
In order to find cosmological signatures of branes, it is important to
study the formation and the evolution of cosmological
defects.\footnote{Inflation in models of 
low fundamental scale   
are discussed in ref.\cite{low_inflation, matsuda_nontach,
matsuda_defectinfla}.
Scenarios of baryogenesis in such models are discussed in ref.
\cite{low_baryo, Defect-baryo-largeextra, 
Defect-baryo-4D}, where defects play distinguishable roles.
The curvatons might play significant roles in these
models\cite{matsuda_curvaton}. 
Moreover, in ref.\cite{topologicalcurvaton} it has been discussed that
topological defects can play the role of the curvatons.}
Defects in brane models such as monopoles, strings, domain walls and
Q-balls are discussed in ref.\cite{matsuda_necklace, BraneQball,
matsuda_monopoles_and_walls, incidental, matsuda_angleddefect,
overproduction}, where it has been  
concluded that not only strings but also other defects can
appear.\footnote{See fig.\ref{fig:history}}

\begin{figure}[ht]
 \begin{center}
\begin{picture}(410,340)(0,0)
\resizebox{13cm}{!}{\includegraphics{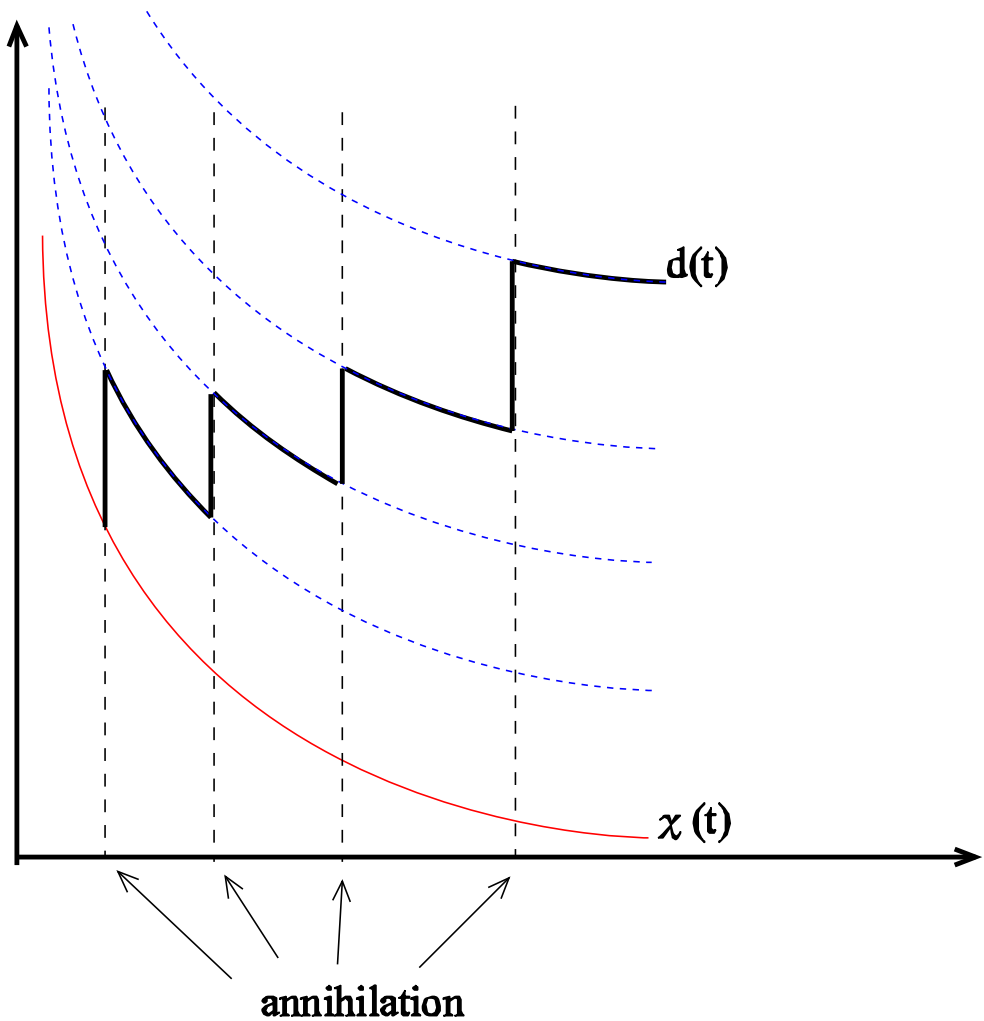}} 
\end{picture}
\caption{The evolution of the necklaces between each annihilation is
  well described by the equation (\ref{req}).
  During the period between each annihilation, the evolution of $d$ is
  therefore given by $d\propto t^{-\kappa} \sim t^{-1}$.
  Assuming that $\kappa$ is a constant during the evolution, 
  one can understand that $\chi$ is a continuous decreasing function
  while the practical value of $d$ is discontinuous at each
  annihilation. }  
\label{fig:annihilation}
 \end{center}
\end{figure}

The purpose of this paper is to find distinguishable properties of
the cosmic necklaces.
We focus our attention on dark matter production from loops of 
cosmic necklaces.\footnote{See also ref.\cite{DMfromStrings}}
The evolution of the networks of cosmic necklaces is first discussed by
Berezinsky and Vilenkin\cite{vilenkin-necklace}.
If one started with a low density of monopoles,
one can approximate the evolution of the system by the standard
evolution of a string network.\footnote{In this paper, we use a
dimensionless parameter $r$, which denotes  
the ratio of the monopole energy density to the string energy density
per unit length. The low density of monopoles corresponds to $r \ll 1$.}
Then the authors found that if one could ignore monopole-antimonopole
annihilation, the 
density of monopoles on strings would increase until the point where
the conventional-string approximation breaks down.
However, in ref.\cite{vilenkin-necklace} the authors leave the detailed
analysis of the evolution of such systems to numerical simulations, in
particular the effect of monopole-antimonopole annihilation.
Later in ref.\cite{necklace-simulation}, numerical simulations of cosmic
necklaces are performed, in which it has been found that the string motion
is periodic when the total monopole energy is much smaller than the 
string energy, and that the monopoles travel along the string and
annihilate with each other.
In this paper, based on the results obtained in
ref.\cite{necklace-simulation}, we will assume that
monopole-antimonopole annihilation is an efficient process.
We can therefore approximate the evolution of the necklaces by the
standard evolution given in ref.\cite{vilenkin-necklace}, at least
 during the period between each annihilation.\footnote{See
fig.1 in ref.\cite{matsuda_pbh} and fig.\ref{fig:annihilation}.}

Let us consider the necklaces produced after or at the end of
inflation\cite{matsuda_necklace}. 
In the case that the compact manifold is not simply connected,
there is a possibility that loops wind around nontrivial
circles.\footnote{See the figure in ref.\cite{matsuda_necklace}}
If such winding states are stabilized, the
simple statistical argument of a random walk indicates that 
the winding states on a long string loop should survive.
If loops were chopped off from such strings, heavy winding states (coils
or 
cycloops) would remain\cite{matsuda_necklace, cycloops}.
In this case, the nucleating rate of the winding states decreases,
while the mass increases with time\cite{matsuda_pbh}.
For example, one may simply assume that $\chi$
always ``increases'' with time due to the conventional expansion of the
Universe.
In this case, the evolution of the scale factor of the distance between
kinks $\chi(t)$, which is the step length between each random walk, 
is not assumed to be affected by the string dynamics.
Then, one can easily find that the mass depends on time as 
$m_{coil} \propto t^{1/4}$ during the radiation dominated
epoch\cite{matsuda_pbh}. 
A similar argument has been discussed in ref.\cite{cycloops}, however
the authors have assumed that $\chi$ is a
constant that does not depend on time.
On the basis of this assumption, it has been concluded that the mass of
cycloops depends on time as $m_{cycloops} \propto t^{1/2}$. 
However, considering the result obtained in
ref.\cite{vilenkin-necklace}, it is obvious that one cannot simply 
ignore the possibility that $\chi$ ``decreases'' with time.
Obviously, one cannot ignore this possibility even in the cases
where the actual distance between monopoles increases with
time.\footnote{In any case, $\chi$ is always much smaller than the
actual distance between monopoles. See fig.\ref{fig:annihilation}.} 
In this paper, we assume that $\chi$ 
depends on time as
\begin{equation}
\label{dk}
\chi(t) \propto t^{k-1},
\end{equation}
where $k\simeq 0$ corresponds to the ``natural'' solution 
$\kappa_g-\kappa_s\simeq 1$ in
ref.\cite{vilenkin-necklace}, and $k=1$ corresponds to the assumption
that was made in ref.\cite{cycloops}.
\footnote{$k=3/2$ corresponds to the ``simplest''(but not reliable due
to the string dynamics) limit that we have mentioned above.
Considering the result obtained in ref.\cite{vilenkin-necklace}, 
it should be fair to say that $k\ge 1$ is unlikely in our setups.}
Then the typical mass of the winding state becomes
\begin{equation}
m_{coil} \sim \left(\frac{l(t)}{\chi(t)}\right)^{1/2} m,
\end{equation}
where $l(t)$ is the length of a loop that is chopped off from the long
strings at $t$, and $m$ is the mass of a
monopole.\footnote{Here we should note that;
\begin{enumerate}
\item The necklace is similar to the standard strings when $r < 1$.
\item During the evolution, it is known that the networks of the strings
   emit loops. The typical size of the chopped strings(loops) does
   depend on time.
\item Then, the typical mass of the chopped loops is determined by $r(t)$
   and the length size of the loops. The mass of the coils must depend
   on time. 
\item The winding number of the loops is conserved once it is chopped off
   from the long strings.
\end{enumerate}
}

It has been claimed in ref.\cite{cycloops} that cycloops poses a
potential monopole problem 
because loops wind around a nontrivial circle behave like heavy matter
at radiation epoch. 
Then they have discussed that in order to avoid cycloop domination the
strings must 
satisfy the severe constraint $G\mu < 10^{-14}$.
In their scenario, however, the authors assumed that cosmic strings can
move freely in extra dimensions when dark matter is produced from their
loops, and claimed that the mass of cycloops
increases with time as $m_{cycloops}\propto t^{1/2}$ if the strings obey
statistical model of random walk, and that $m_{cycloops}\propto t$
if velocity correlation is considered. 
Obviously, their assumption of the free motion depends on the potential
that lifts the moduli that parameterizes extra dimensions.
Cycloops turn into necklaces when the ``lift'' becomes significant.
Moreover, as we have stated above, the evolution of $m(t)$
depends crucially on the value of $k$ in eq.(\ref{dk}). 
Therefore, the result must be reexamined if the ``lift'' becomes
significant before the dark-matter production, or the deviation from 
$k=1$ becomes crucial.
Moreover, if one wants to examine the relic density of
superheavy dark matter that is produced from string loops, one cannot
simply ignore frictional forces from the thermal plasma, which
in some cases determines the 
string motion in the early Universe, as was already discussed for 
vortons\cite{vorton_branden}.
Therefore, it should be important to consider string networks at the
damped epoch if one wants to calculate the density of relic superheavy
dark matter, such as coils and cycloops.
Of course, one cannot simply ignore the effective potential that lifts
the moduli, particularly the one lifts the flat direction that
corresponds to the string motion in the compactified space.\footnote{The
effective potential is supposed to be significant at $H \sim m_{moduli}$.}
If the potential stabilizes the vacuum at $t=t_{p}$, after this time 
one should consider cosmic necklaces/coils instead of cycloops.
For example, if the potential is lifted by the effect of supersymmetry
breaking, one can assume $t_p \sim m_{3/2}^{-1}$, where $m_{3/2}$ is
the mass of the gravitino.
Alternatively, in the case that the stabilization is induced by brane
dynamics, one can assume that the stabilization occurs just after
the string formation.
Even in this case, it is natural to consider random distribution of the
``vacua'' on the cosmic strings if brane annihilation or brane
collision is so energetic that the strings have enough kinetic energy to
climb up the potential hill at least just after they are produced.
Moreover, in more generic models of necklaces, one can assume that
monopoles are produced {\bf before} string formation.

To understand the stability of the winding states,
we consider necklaces whose loops are stabilized by their windings.
The winding state could be a higher-dimensional object(brane) that winds
around a nontrivial circle in the compactified space, or could be
a nonabelian string whose loop in the moduli space 
is stabilized by the potential barrier.
In our previous paper\cite{matsuda_pbh} we have considered two concrete
examples for the winding state. 
The first  is an example of cosmic strings produced after brane
inflation, and the second  is an example of nonabelian necklaces.
Of course one can reconstruct the nonabelian necklaces by using the
brane language.
We will comment on this issue in appendix to clarify the origin of the
frictional force acting on the necklaces.

In this paper, we consider dark matter(DM) production at the damped
epoch.
We have obtained distinguishable properties of
the networks of cosmic necklaces.

\section{Dark matter production from cosmic necklaces}
\subsection{Mass of the stable winding states}
As we have discussed in our previous paper\cite{matsuda_pbh}, it is
appropriate to consider necklaces of $r \ll 1$ because of  
the efficient annihilation of monopole-antimonopole.
Then, the simple  
statistical argument of random walk indicates that about $n^{1/2}$
of the initial $n$ monopoles on a long string could survive.
Let us make a brief review of the resuld obtained in
ref.\cite{matsuda_pbh}. 

Here the important quantity for the necklace evolution is the
dimensionless 
ratio $r=m/\mu d$.
During the period between each annihilation, one can follow the
discussions given in ref.\cite{vilenkin-necklace}. 
The equation for the evolution of $r$ is 
\begin{equation}
\label{req}
\frac{\dot{r}}{r}=-\kappa_s t^{-1} + \kappa_g t^{-1},
\end{equation}
where the first term on the right-hand side describes the string
stretching that is due to the expansion of the Universe, while the second
describes the effect of string shrinking due to gravitational
radiation.
Using the standard value from string simulations, it has been concluded
that the reasonable assumption is $\kappa_g > \kappa_s$.
The solution of eq.(\ref{req})is
\begin{equation}
r(t) \propto t^{\kappa}.
\end{equation}
Considering the order-of-magnitude estimation, 
one can obtain $\kappa = \kappa_g-\kappa_s \sim
1$\cite{vilenkin-necklace}. 
Therefore, disregarding monopole-antimonopole annihilation, 
one can understand that the distance between monopoles ``decreases'' as
$d \propto t^{-\kappa} \sim t^{-1}$ until the conventional-string
approximation is broken by dense monopoles.
In our case, it should be reasonable to think that the ``distance
between monopoles'' 
obtained above is corresponding to $\chi$, the step length between each
random walk\footnote{See fig.1 in ref.\cite{matsuda_pbh}.}.
To be precise, if $n$ monopoles are obtained disregarding
monopole-antimonopole annihilation, one should obtain $n^{1/2}$
monopoles after annihilation.
If the annihilation is an efficient process, $\chi$
 can continue to decrease while the actual value of $r$ remains
small.\footnote{See fig.\ref{fig:annihilation}.}
In some cases, assuming efficient annihilation,
the actual number density of monopole may become a
constant.\cite{matsuda_pbh}\footnote{It has been 
suggested in ref.\cite{vilenkin-necklace} that 
 the ratio $r$ might have an attractor point. 
Our result obtained in
 ref.\cite{matsuda_pbh} supports this conjecture.}

Then it is easy to obtain the mass of the stable
relic\cite{matsuda_pbh}; 
\begin{equation}
M_{coil}(t) \sim n(t)^{1/2} m.
\end{equation}
Here the number of monopoles that are ``initially''
contained in a loop is given by
\begin{equation}
\label{number-monopole_ini}
n(t) \sim \frac{l(t)}{d(t_n)\times \left(\frac{t}{t_n}\right)^{k-1}}.
\end{equation}
In the scaling epoch one can obtain $M_{coil} \propto t$ for $k=0
(\kappa = 1)$,
which is similar to the result obtained from velocity
correlation\cite{cycloops} despite the 
fact that we are not assuming free motion in extra dimensions.

\subsection{Frictional force acting on strings and monopoles}
A monopole moving through plasma in the early Universe experiences a
frictional force due to its interaction with charged particles.
The gauge and Higgs fields which have nonzero vacuum expectation value
inside strings or monopoles can couple to various other fields.
This results in effective interactions between the defects and the
corresponding light particles.
Naively, one might expect the typical length scale of the scattering
cross-section to be comparable to the physical thickness of the defects,
$\delta_s$ or $\delta_m$.
Of course, this expectation is incorrect.
The actual cross-section is determined by the wavelength of the incident
particle, which means that the scattering cross-section becomes 
$\sigma_s \sim T^{-1}$ for strings and $\sigma_m \sim T^{-2}$ for
monopoles.
A rough estimate of this force is\cite{vilenkin_book}
\begin{equation}
\label{frictional_monopole}
F_{m0}= \beta_m T^2 v,
\end{equation}
where $\beta_m$ is a numerical factor.
Here we have assumed that a monopole is moving with a nonrelativistic
velocity, $v$.
A string moving through plasma in the early Universe experiences a
frictional force from the background plasma.
The force per unit length is\cite{vilenkin_book}
\begin{equation}
\label{frictional_string}
F_s = \beta_s T^3  v,
\end{equation}
where $\beta_s$ is a numerical factor or order unity.
The frictional  force becomes negligible at 
\begin{equation}
t_* \sim (G\mu)^{-1}t_s,
\end{equation}
where $t_s$ is the time of string formation.

Although the standard result that we have obtained above describes the
essential properties of the necklaces, one should consider rather
peculiar situations when the number of species of the monopoles becomes
large, $N_n \gg 1$.
In appendix A, we show how monopoles affect the frictional force acting on
necklaces with $N_n \gg 1$.

\subsection{Typical curvature radius $R$ and typical distance $L$ at the
  damped epoch.}
Since we are considering the efficient annihilation of
monopole-antimonopole on the necklaces, we can assume $r\ll 1$.
Then the frictional force acting on necklaces is given by
(\ref{frictional_string}).\footnote{Here we disregard the ``peculiar''
properties of the necklaces that will be discussed in appendix A.}
The damped epoch corresponds to the highest string densities and so
should be important for baryogenesis, vorton
formation\cite{vorton_branden} and other effects.
In this paper, we consider production of superheavy states(coils) during
this epoch and examine cosmological constraints.
To start with, let us calculate the characteristic damping time for the
necklaces. 
Denoting the kinetic energy per unit length and energy dispersion ratio 
by$\epsilon$ and $\dot{\epsilon}$, the characteristic damping
time becomes
\begin{eqnarray}
t_d &\sim& \frac{\epsilon}{\dot{\epsilon}}\nonumber\\
&\sim& \frac{(r+1)\mu v^2}{F_s v}\nonumber\\
& \sim &
\frac{\mu}{T^3} \times \frac{r+1}{\beta_s}.
\end{eqnarray}
Note that $t_d$ is much smaller than the Hubble time, $t\sim
M_p/T^2$. 

Now we can calculate the typical curvature radius $R(t)$ and 
the typical distance between the nearest string
segments in the network, $L(t)$.
The force induced by the tension of a string of curvature radius $R$ is
\begin{equation}
F_t \sim \mu/R.
\end{equation}
It is easy to find the approximate value of the corresponding
acceleration,
\begin{equation}
a_t \sim \frac{F_t}{(r+1)\mu}\sim \frac{1}{(r+1)R}.
\end{equation}
At the damped epoch the string can be accelerated only for a time
period $\sim t_d$, which suggests that the string moves with the
typical velocity
\begin{equation}
v\sim a_t t_d.
\end{equation}
After the time period $t_d$, the force induced by the string tension is
balanced by 
the frictional force from the background plasma.
The balancing speed is obtained from the condition for the force balance;
\begin{equation}
\label{friction_tension}
\frac{\mu}{R}\sim F_s.
\end{equation}
The typical curvature radius will grow as $R(t)\sim
v t \sim a_t t_d t$\cite{vilenkin_book}.
We can therefore obtain the result 
\begin{equation}
R(t)\sim \sqrt{\frac{t_d t}{r+1}}.
\end{equation}
One can assume that $R(t)$ is as the same order as the typical coherence
length $\xi$;
\begin{equation}
\label{equationforxi}
\xi(t) \sim R(t)\sim
\left(\frac{\mu M_p}{T^5 \beta_s}\right)^{1/2},
\end{equation}
which is always much smaller than the horizon size.
As in the conventional scenario of the string network evolution,
$R(t)$ depends on the Hubble time as $R(t)\sim t^{5/4}$.
In this case, $R(t)$ grows faster than the horizon scale $t$.
Therefore, as long as the evolution of the networks of necklaces is
approximated by the evolution of conventional string networks, 
small-scale irregularities and loops of 
size smaller than $R$ should be damped out in less than a Hubble time.
Since smaller wiggles are suppressed at the damped epoch,
the typical size of the loops is $l(t)\sim R(t)$\cite{vorton_branden}.

\subsection{Loops at the damped epoch}
Our next task is to calculate the typical number density of the loops.
In this case, we should take into account the low reconnection rate of
the necklaces, $p \ll 1$.
As we have discussed above, a necklace that has $N_n$
degenerated vacua on its worldvolume is macroscopically the
same as  a string with low reconnection rate $p \sim N_n^{-1}$.
In this case, in order to have one reconnection per Hubble time, a
necklace needs to have $\sim p^{-1}$ intersection per Hubble time.
Then the number of such necklaces per volume $(v t)^3$ is $\sim p^{-1}$.
Therefore, the mean number density of string loops
becomes\cite{vorton_branden}
\begin{equation}
\label{number-density}
n_l \sim \frac{1}{p \xi^3}.
\end{equation}
For the loops to be stabilized, they must wind more than one time around
the nontrivial circle in their moduli space.
Therefore, the loops are stabilized when loops of
the length $l(t)$ contain at least $N_n$ monopoles after 
monopole-antimonopole annihilation.
Then the condition for the stabilization becomes
\begin{equation}
\label{stabilicon}
l(t) > \frac{m N_n}{\mu r}.
\end{equation}
Since the typical length of the chopped loops grows with time, 
we may assume that the production of the stable loops starts at $t_f$.
Then the mass of the coils that are produced from loops becomes
\begin{equation}
\label{massofcoils}
m_{coil}(t_f)\sim \mu r(t_f) l(t_f).
\end{equation}
From eq.(\ref{number-density}) and (\ref{massofcoils}) one can obtain
the energy density
\begin{equation}
\rho_{coil}(t_f)\sim \frac{\mu r(t_f) l(t_f)}{p \xi(t_f)^3}\sim
\frac{T^5_f \beta_s r}{p  M_p}.
\end{equation}
Since the coils behave as nonrelativistic matter, the evolution of
their number density becomes
\begin{equation}
n_{coil}(T) \sim \frac{1}{p} 
\left(\frac{T_f^3 \beta_s}{\mu M_p} \right)^{3/2} T^3.
\end{equation}
The situation that we are considering in this paper is quite similar to
the one that have been considered to obtain the vorton relic density.
Therefore, following the analysis in ref.\cite{vorton_branden}, it 
is easy to obtain the result
\begin{equation}
\rho_{coil}(T)\sim \frac{T^2_f \beta_s r T^3}{p  M_p }.
\end{equation}
Obviously, the significant property of the above result is the
disappearance of the string tension, $\mu$.
Although the density $\rho_{coil}$ still depends on the formation time
$t_f$, we should remember that $t_f$ is determined by the stabilization
condition (\ref{stabilicon}).
As $t_f$ depends strongly on the initial value of $r$
and the numerical constant $k$ that controls the evolution of $r$,
the cosmological constraints  cannot put the direct bound on $\mu$
and $m$.

\subsection{Cosmological constraints}
Perhaps the most robust prediction of the standard cosmological particle
model is the abundance of the light elements that were produced during
primordial nucleosynthesis.
Nucleosynthesis occurs at the temperature $T_N \sim 10^{-4} GeV$.
In order to preserve the well-established scenario of nucleosynthesis,
it is needed that the coil distribution should satisfy 
$\rho_{coil} < T_N^4$.\footnote{In this paper, we have neglected the
changes in the particle-number weighting factor $g^*$ in the temperature
range under consideration.} 
In our case, the condition becomes
\begin{equation}
T_f < 10^{7}GeV\times \left[\frac{p}{10^{-2}}\right]^{1/2}
\left[\frac{1}{\beta_s}\right]^{1/2}
\left[\frac{10^{-3}}{r}\right]^{1/2}.
\end{equation}
$\beta_s$ is a numerical factor of order unity\cite{vilenkin_book}.
The typical value of $p$ is $1>p>10^{-2}$\cite{previous-onlystrings}.
In the case that the number of the windings per loop is proportional to
the length of the loop, $r$ is a constant.
In this sense, $r$ has a fixed point in the scaling epoch, if the
evolution of the necklaces is determined by the standard
equation\footnote{See ref.\cite{matsuda_pbh, vilenkin-necklace} and
fig.\ref{fig:annihilation}}.  
On the other hand, one can obtain $r \propto t^{-1/8}$ from
(\ref{equationforxi}) and (\ref{windingnumber}), which suggests that $r$
is a slowly varying function during the damped epoch.
The initial value of $r=m/\mu d$ is obtained if one assumes that 
initially $d$ is as large as the Hubble radius $H^{-1}\sim M_p/\mu$.
Then, one can obtain $r_0 \sim m/M_p$.
For $m \sim M_{GUT} \sim 10^{16}GeV$, $r_0$
becomes $r_0 \sim 10^{-3}$.\footnote{The mass of the monopoles on the
necklaces depends on the structure of the internal space, which is
highly model-dependent. If it winds around large extra dimension, its
mass becomes huge even if the fundamental scale is as low as O(TeV).}

We now consider a stronger constraint.
The stronger constraint is obtained if the winding
states are sufficiently stable and can survive until the present
epoch.\footnote{Coils and cycloops that wind around a
nontrivial circle in the compactified space are stable
due to the topological reason.
However, coils that are stabilized due to the potential barrier
may decay by tunneling. 
The lifetime of such unstable coils is determined by the potential that
lifts the moduli.
The peculiar cases of the unstable coils are interesting but highly
model-dependent.}  
Following ref.\cite{vorton_branden}, one can easily obtain
\begin{equation}
T_f < 10^{5}GeV\times \left[\frac{p}{10^{-2}}\right]^{1/2}
\left[\frac{1}{\beta_s}\right]^{1/2}
\left[\frac{10^{-3}}{r}\right]^{1/2}.
\end{equation}
The above constraint seems already quite stringent.
However, here we examine the above condition in more
detail. 
Let us consider the case where $\chi$ evolves as $\chi \propto t^{-1}$. 
This assumption is appropriate both for the strings in free
motion(velocity correlation has been discussed in
ref.\cite{cycloops}) and the necklaces(see fig.\ref{fig:annihilation}).
Then the expected winding number per loop $<n>$ is given by
\begin{equation}
\label{windingnumber}
<n>\sim \sqrt{\frac{\xi}{\chi}}\sim 
\sqrt{\frac{\mu^{1/2}M_p^{1/2}T^{-5/2}\beta_s^{1/2}}{\chi_0 \times
\frac{T^2}{\mu}}}
\end{equation}
where $\mu^{-1/2} < \chi_0 < M_p/\mu$ is the initial length of $\chi$
when strings are formed.\footnote{$\chi_0 \sim \mu^{-1/2}$ is used in
ref.\cite{cycloops}.}

Having the modest assumption that $\chi_0 \sim M_p/\mu$,
one can obtain the temperature $T_f$ from 
the equation $<n(T_f)>\sim 1$,
\begin{equation}
\label{tf}
T_f \sim \mu^{1/2} \left(\frac{\mu^{1/2}}{M_p}\right)^{1/9},
\end{equation}
which suggests that the dark-matter production starts soon after the
string formation.
One can therefore understand that the effect of a small
deviation from $k=0$ is not significant for the obtaind bound.
Even in this case (where we have the modest assumption $\chi_0 \sim
M_p/\mu$), the upper bound for $G\mu$ is about
\begin{equation}
\label{finalresult}
G\mu< 10^{-23}\times \left[\frac{p}{10^{-2}}\right]^{9/10}
\left[\frac{1}{\beta_s}\right]^{9/10}
\left[\frac{10^{-3}}{r}\right]^{9/10}.
\end{equation}
Therefore, our result (\ref{finalresult}) puts a severe bound on the
inner structure of brane models, in the case that stable coils are
produced.

As we have mentioned in the previous section, the significant point
is that the string tension $\mu$ has been disappeared from the
cosmological constraint. 
The obtained bound is for $T_f$, as we have discussed above.
Of course, one may think that the bound is not significant because $T_f$
is determined by the dynamics of the cosmic necklaces.
It is true that $T_f$ seems to depend crucially on the initial
configuration and the numerical constant $k$ that controls the evolution
of $r$.
However, even in the case where the initial distance between
monopoles is as large as the Hubble radius, and the evolution of the
necklaces is determined by the standard equation (\ref{req}),
the bound we have obtained is eq.(\ref{finalresult}), which is of
course quite stringent.
Moreover, the effect of a small deviation from $k=0$ is not significant
for the obtaind bound, as we have discussed above.

\begin{figure}[ht]
 \begin{center}
\begin{picture}(500,300)(0,0)
\resizebox{17cm}{!}{\includegraphics{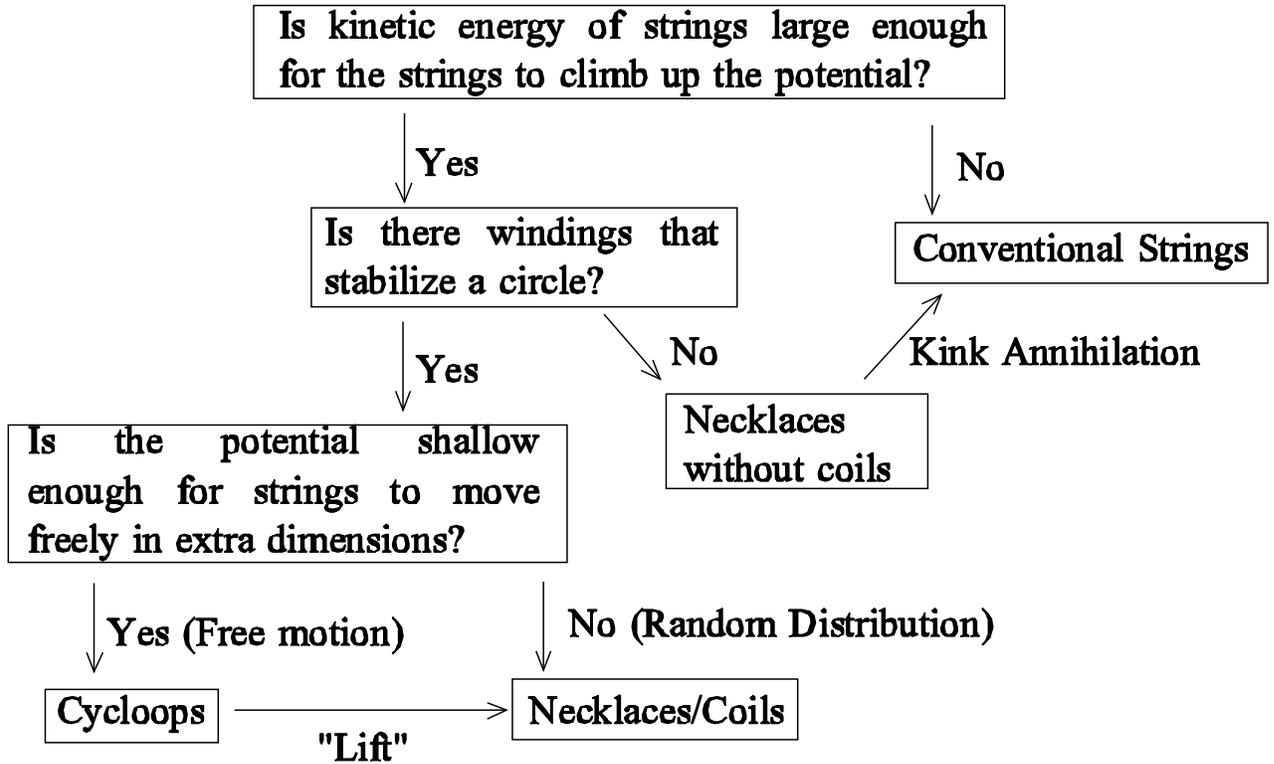}} 
\end{picture}
\caption{This picture shows typical situations when the networks of
  necklaces and coils become significant. 
  It is important to note that cycloops turn into necklaces/coils 
  when their free motion in extra dimensions is stopped by the
  potential.
  In this sense, late-time production of PBH relics must be investigated
  in the framework of necklaces and coils\cite{matsuda_pbh}.}
\label{fig:idea1}
 \end{center}
\end{figure}

Here we should make some comments about the discrepancy
between our result and the result obtained for cycloops in
ref.\cite{cycloops}\footnote{See also Fig.\ref{fig:idea1}}. 
In ref.\cite{cycloops}, it has been claimed that cycloops poses a
potential monopole problem because such loops behave like heavy matter
at radiation epoch. 
Then they have shown that in order to avoid cycloop domination the
strings must satisfy the severe constraint $G\mu < 10^{-14}$.
However, in their analysis they have disregarded the damped epoch and 
also made a nontrivial assumption that the strings move freely in the
internal space when the significant amount of the winding state
is produced.
In general, the damping term becomes negligible at
temperatures\cite{vilenkin_book} 
\begin{equation}
T \ll T_* = G\mu M_p,
\end{equation}
which is always lower than $T_f$ that we have obtained in (\ref{tf}).
Therefore, it is appropriate to consider the production of dark
matter in the damped epoch rather than in the scaling epoch.

\section{Conclusions and Discussions}
Cosmic strings have recently gained a great interest because they are
formed in a large class of brane inflationary models.
The most interesting story would be that cosmic strings in brane
models are distinguished from conventional 
cosmic strings in future cosmological observations. 
It has already been discussed that such strings may
lead to observational predictions that can be used to distinguish 
brane world from conventional phenomenological
models\cite{vilenkin-lowp, matsuda_necklace, matsuda_pbh}.
If the strings in brane models are branes that can move along
compactified space, and also if there are degenerated vacua
along the compactified space, the strings turn into necklaces.
Moreover, in the case that the compact manifold in not simply connected,
a string loop that winds around a nontrivial circle is stabilized.
Since the existence of the (quasi-)degenerated vacua and the nontrivial 
circle is a common feature of brane models, it should be important to
examine cosmological constraints on cosmic necklaces and their 
stable winding states.
If the existence of a stable winding state is excluded, necklace becomes
a probe of the compactified space.  
In this paper, we have considered the production of dark matter from
loops of cosmic necklaces.
The bounds we have obtained are stringent.
Our result suggests that necklaces may put stringent bound on brane
models, as far as the standard scenario of the necklace evolution is
applicable. 

Finally, we will comment on the cosmological production of winding
states in KKLT models, which seems far from obvious.
In models that have been discussed so far in the literature, 
there is no winding state at the bottom of the inflation throat, because
it is not required for successful inflation.
Moreover, in some cases there is a (possibly large) potential barrier 
that blocks the strings from moving out of the throats to wind around
the bulk. 
In models where the potential barrier is effective, one can see that
nontrivial circle is hidden from the 
strings, which makes the production of winding states negligible.
In this case, our analysis cannot put bounds on the scale.
On the other hand, in the case when the potential barrier cannot block
the string motion, strings can penetrate into the bulk and may produce
winding states.
We will leave the detailed argument about the KKLT models for future
work, since the analyses in the KKLT models are highly model dependent
due to the mechanisms of inflation and reheating that are still
developing.

\section{Acknowledgment}
We wish to thank K.Shima for encouragement, and our colleagues in
Tokyo University for their kind hospitality.

\appendix

\section{Origin of the frictional force}
\begin{figure}[ht]
 \begin{center}
\begin{picture}(500,300)(0,0)
\resizebox{15cm}{!}{\includegraphics{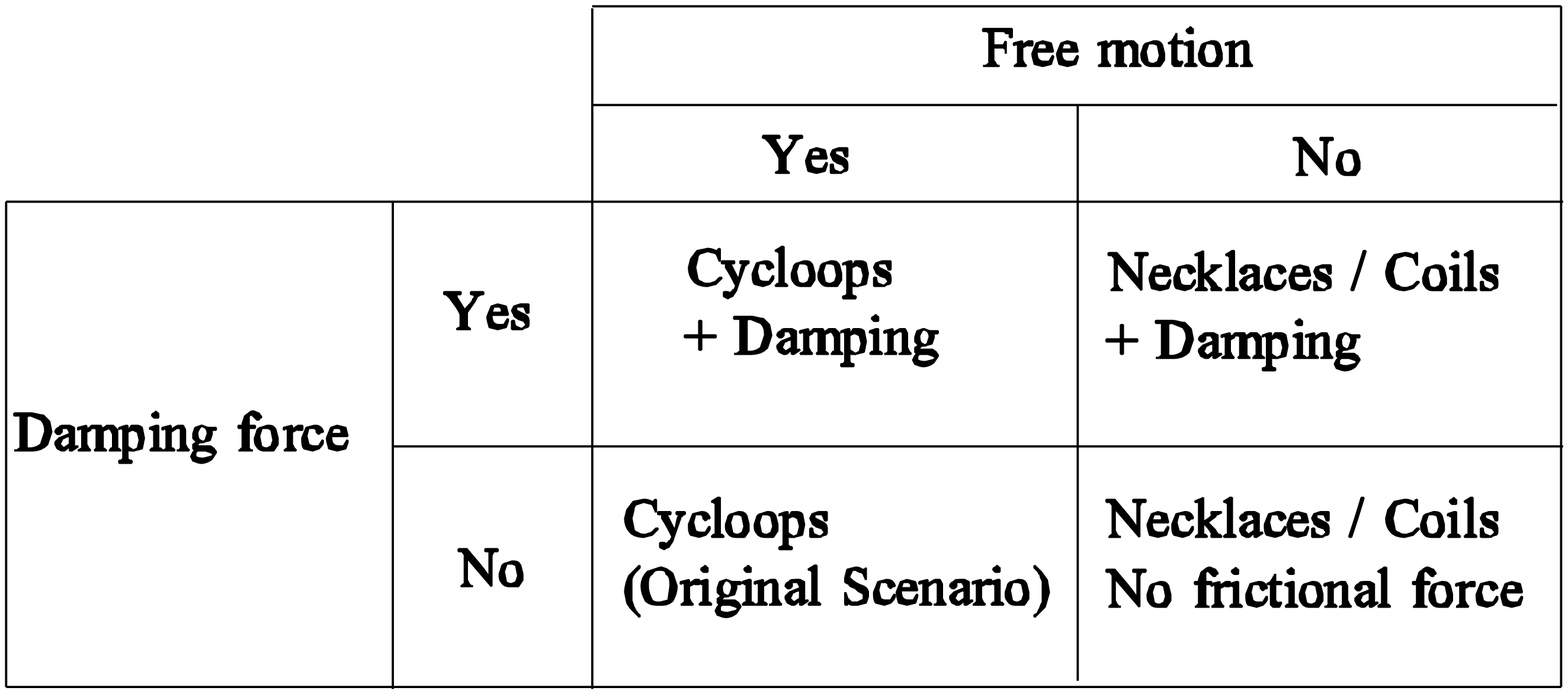}} 
\end{picture}
\caption{Frictional forces become important only when strings can interact
  with thermal plasma. 
  In the case when thermal plasma is localized on a distant brane or a
  distant throat, interactions between cosmic strings and thermal plasma
  are suppressed by exponential factor, thus the frictional forces
  should be negligible.
  In this case the strings must be produced somewhere at a distance
  while reheating occurs on the standard-model brane. 
  One may think that the situation is peculiar, however
  it is actually possible to construct models in which strings
  are located on a hypersurface (or throat) which is far-distant from
  standard-model brane on which reheating is induced after
  inflation\cite{Distant-reheat}. 
  Of course, one cannot ignore the possiblity that reheating in bulk
  fields is not negligible at least just after inflation.
  One may consider another possibility that strings are produced
  after angled inflation\cite{matsuda_angleddefect}.
  In this case, cosmic strings are
  extended between branes, thus they can feel frictional forces from the
  plasma that is localized on the spacetime-filling branes.  
  It should be noted that similar cosmic strings can be produced by
  later thermal phase transition in brane models, if the phase
  transition is explained by brane recombination.}   
\label{fig:idea2}
 \end{center}
\end{figure}

Apart from KKLT multi-throat models\footnote{If one considers KKLT
multi-throat models in which standard model branes 
are localized at the bottom of the SM throat while inflation occurs at
another throat (inflation throat), one may think that thermal plasma is
localized at the bottom of the SM throat while cosmic strings are produced
in the inflation throat.
In this case, it seems obvious that strings cannot feel frictional
forces from the plasma at a distance.
However, what we are considering in this paper is the string networks
just after the string production.
Even if the reheating due to tunneling is successful, the original decay
products just after brane annihilation are produced in the inflation
throat, where the cosmic strings are produced.
In the KKLT scenarios, we need to understand more clearly 
the mechanism of inflation and reheating, which are highly
model-dependent.
Thus, we will leave the detailed argument about the KKLT models for
future work. See also Fig.\ref{fig:idea2}}, 
one may think that the thermal plasma is localized on 
spacetime-filling branes and cannot interact with cosmic strings.
In the case that thermal plasma is localized on a distant brane or a
distant throat, interactions between cosmic strings and thermal plasma
are suppressed by exponential factor, thus the frictional forces
should be negligible.
On the other hand, in this case one should assume that strings are
produced somewhere at a distance while reheating occurs on the
standard-model brane.  
One may think that the situation looks peculiar.
Of course we know that it is possible to construct models in which
reheating seems to occur only for the fields localized on branes.
Thus, to understand the origin of the frictional forces, we think it is
helpful to consider models in which 
the interactions between strings and thermal plasma on the
spacetime-filling branes are obvious.
For the cosmic strings that are produced at the last stage
of brane inflation, we will consider strings produced after angled
inflation.
In angled inflation, cosmic strings are extended between
branes\footnote{See Fig.1 in ref.\cite{matsuda_angleddefect} and Fig.1
in the first paper in ref.\cite{matsuda_monopoles_and_walls}.}, 
thus they can feel frictional forces from the plasma on the
spacetime-filling branes.  
The same kind of cosmic strings can be produced
at later thermal phase transition, if the phase
transition is accompanied by brane
recombination\cite{matsuda_monopoles_and_walls}. 

In the case that the lift of the potential is not important, one may
use cycloops to obtain DM abundance.
Our analysis on DM production is still useful in this case,
if there are interactions between plasma.
One can calculate DM production from cycloops in damped epoch,
which is consistent with our result because the evolution of the mass of
the cycloops is given by $m(t)\propto t$\cite{cycloops}.

There may be models in which strings are produced
at a distance from the standard-model branes.
In these models, damped epoch is highly model-dependent and not obvious.

\section{Nonabelian strings in brane construction}
As we have discussed in this paper, we think it is natural to consider
necklaces in brane models.
On the other hand, more explanations should be needed to understand
whether one can construct necklaces and coils in four-dimensional gauge
theory.
To construct necklaces in four-dimensional gauge theory, 
the moduli that parameterizes the string motion in extra dimensions must
be 
replaced by a flat direction that appears in the two-dimensional
effective action on the strings.
Let us consider the dynamics of cosmic strings living in a nonabelian
$U(N_c)$ gauge theory that is coupled to $N_f$ scalar fields $q_i$,
which transform in the fundamental representation\cite{tong_hashimoto};
\begin{equation}
L = \frac{1}{4e^2}Tr F_{\mu\nu}F^{\mu\nu} + \sum^{N_f}_{i=1}
{\cal D}_{\mu}q_i^{\dagger}{\cal D}_{\mu}q_i^{\dagger}
-\frac{\lambda e^2}{2}\left(\sum^{N_f}_{i=1} q_i \otimes q_i^{\dagger}
-v^2\right)^2.
\end{equation}
One can see that $SU(N_f)$ flavor symmetry appears in
the above Lagrangian rotates the scalars. 
Then, it is possible to include explicit symmetry breaking terms into the
Lagrangian, which breaks global flavor symmetry.
The most obvious example is a small mass term for the scalars;
\begin{equation}
V_{br1}\sim \sum_{i}m_i^2 q_i^{\dagger}q_i,
\end{equation}
which shifts the vacuum expectation value to\cite{tong_hashimoto}
\begin{equation}
q^a_i = \left(v^2-\frac{m^2_i}{\lambda e^2}\right)^{1/2}\delta^a_i.
\end{equation}
Then, an abelian vortex in the i-th $U(1)$ subgroup
of $U(N_c)$ can be embedded, whose tension becomes $T_i \sim
\left(v^2-\frac{m^2_i}{\lambda e^2}\right)^{1/2}$.

One may extend the above model to $N=2$ supersymmetric QCD 
or simply include an additional adjoint scalar field $\phi$.
Then, the typical potential for the adjoint scalar is given by\cite{tong_hashimoto} 
\begin{equation}
V_{br2}\sim \sum_{i}q_i^\dagger(\phi-m_i)^2 q_i.
\end{equation}
Be sure that the potential breaks $U(1)_R$ symmetry, and
the tensions of the strings degenerate in this case.

Alternatively, one can consider supersymmetry-breaking potential
that could be induced by higher-dimensional effects,
\begin{equation}
V_{br3}\sim \sum_{i}q_i^\dagger(|\phi|^2-m^2) q_i,
\end{equation}
which preserves $U(1)_R$ symmetry.
In this case, due to D-flatness condition if supersymmetry is imposed,
the vacuum expectation value of the adjoint field is placed on a circle
and given by\cite{matsuda_monopoles_and_walls}
\begin{equation}
\phi=m\times  
diag(1,e^{\frac{2\pi}{N_c}}, e^{\frac{2\pi}{N_c}\times 2},
...e^{\frac{2\pi}{N_c}\times (N_c-1)}).
\end{equation}
One can break the remaining classical $U(1)_R$ symmetry by adding an
explicit breaking term, or by anomaly\cite{matsuda_monopoles_and_walls,
tong_hashimoto}.  

In any case,  strings living in different $U(1)$ subgroups can transmute
each other by kinks(walls on their 2D worldvolume) that interpolate
between (quasi-)degenerated vacua. 

Now our main concern is whether it is possible to construct ``winding''
states from nonabelian necklaces, which look like coils in 
brane models.
A similar argument has already been discussed by Dvali, Tavartkiladze and
Nanobashvili\cite{winding_wall} for $Z_2$ domain wall in
four-dimensional theory.
The authors have discussed that similar ``windings'' may stabilize the
wall-antiwall bound state if the potential is steep in the radial
direction.
Of course, one can apply similar argument to nonabelian necklaces.
In our case, windings can be stabilized if (for example) the origin is
lifted by an effective potential $\sim \phi^{-n}$. 
The important point here is whether the absolute value of the
scalar field can vanish inside the bound state of walls(kinks).
If  ``windings'' of such kinks cannot be resolved due to the potential
barrier near the origin, naive annihilation process is inhibited and
stable bound state will remain.

\begin{figure}[ht]
 \begin{center}
\begin{picture}(500,300)(0,0)
\resizebox{16cm}{!}{\includegraphics{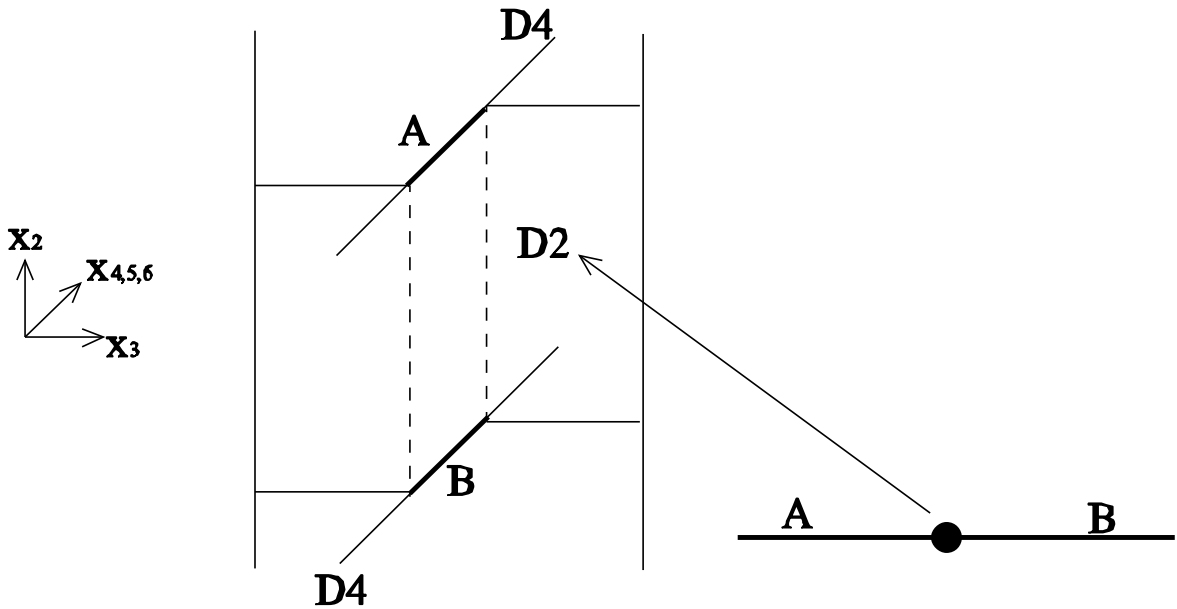}} 
\end{picture}
\caption{This picture shows a typical brane construction of nonabelian
  strings. 
  String parts A and B in the right figure correspond to D2 on A and D2
  on B.
  A kink that interpolates between A and B is a monopole on
  the necklaces.
  If one compactifies the $x^2$ direction on $S^1$ with radius $\rho$,
  one can take T-duality\cite{tit-new}.
  In this case, one can consider motion in the $x^2$ direction, which
  induces windings that are required to stabilize loops of the
  necklaces.
  In the four-dimensional effective action, effective potential for
  $\rho$ will have a high barrier near the origin as expected in
  ref.\cite{matsuda_pbh}.}    
\label{fig:nonabelian}
 \end{center}
\end{figure}

Of course, it is straightforward to construct brane counterpart of the
nonabelian necklaces\cite{tong_hashimoto}.
A typical brane construction is given in fig.{\ref{fig:nonabelian}}.

\section{More on frictional forces acting on necklaces}
As far as the coefficient $\beta_m$ does not much exceed $\beta_s$,
the drag force acting on monopoles does not play crucial role.
In general, the frictional force acting on necklaces is comparable to 
(\ref{frictional_string}). 
However, in the case that the magnetic charge of neighboring monopoles 
is originated from different $U(1)$'s, the frictional force acting on
monopoles is induced by different species of particles.
Then one should sum up the frictional force acting on monopoles.
In this case, one should consider three phases that correspond to the
situations; 
\begin{enumerate}
\item The frictional force acting on necklaces is given by the formula
      (\ref{frictional_string}). In this phase, one can neglect drag
      force acting on monopoles.
\item The frictional  force is dominated by the drag force acting on
      monopoles, but the force is still not saturated.
\item The monopoles become dense and the cross-section of the monopoles
      of the same kind begins to overlap. 
      In this case, the frictional force is saturated and looks
      qualitatively the same as 
      (\ref{frictional_string}).
\end{enumerate}

Let us first consider the boundary between the phases 1. and 2.
As we have stated above, the frictional force acting on strings is
given by eq.(\ref{frictional_string}).
The drag force per unit length, which is  induced by the frictional
force acting on the monopoles is
\begin{equation}
\label{frictional_monopole2}
F_{m}=F_{m0}\times N_m = \beta_m T^2 v N_m,
\end{equation}
where $N_m$ is the number density of the monopoles per unit length,
\begin{equation}
N_m=\frac{\mu r}{m}.
\end{equation}
Therefore, the condition $F_s < F_m$ that is required for the monopoles
to dominate the frictional force becomes
\begin{equation}
T < T_{12} \equiv \frac{\mu r }{m}\frac{\beta_m}{\beta_s}.
\end{equation}
Here $T_{12}$ is the boundary between the phases 1. and 2.

Let us consider the boundary between the phases 2. and 3.
The cross-section between the monopoles of the same kind
begins to overlap when
\begin{equation}
d < 1/(T N_n),
\end{equation}
where $N_n$ is the number of species of monopoles.\footnote{The value of
$N_n$ becomes about $N_n \sim 10^{2-3}$ in angled inflation.}
Therefore, the boundary between the phases 2. and 3. is given by
\begin{equation}
T_{23} \equiv \frac{\mu r}{m N_n}.
\end{equation}
In the phase 3., the frictional force is saturated and becomes
\begin{equation}
\label{frictional_necklace}
F_n \sim \beta_m T^3 v N_n,
\end{equation}
which is qualitatively the same as eq.(\ref{frictional_string}).
Obviously, the phase 2. becomes important in models with $N_n \gg 1$.

\end{document}